\nonstopmode

\documentclass[aps,prl,reprint,showpacs,superscriptaddress,floatfix]{revtex4-2}
\usepackage{hyperref}
\usepackage{xcolor, soul}
\sethlcolor{yellow}
\usepackage{graphicx}
\usepackage{natbib}
\usepackage{epstopdf}
\usepackage{caption}
\usepackage[colorinlistoftodos]{todonotes}
\usepackage{gensymb}
\usepackage{amsmath}
\usepackage{breqn}
\usepackage{float}
\def\msz{$m_s=0$ }

\def\mspm{$m_s=\pm1$ }

\begin{document}
\title{Towards high-sensitivity magnetometry with nitrogen vacancy centers in diamond using the singlet infrared absorption}
\author{Ali Tayefeh Younesi}
\affiliation{Max Planck Institute for Polymer Research$,$ Ackermannweg 10$,$ 55128 Mainz$,$ Germany}
\author{Muhib Omar}
\affiliation{Helmholtz-Institut Mainz$,$ GSI Helmholtzzentrum für Schwerionenforschung GmbH$,$ 55128 Mainz$,$ Germany}
\affiliation{Johannes Gutenberg-University Mainz$,$ 55128 Mainz$,$ Germany}
\author{Arne Wickenbrock}
\affiliation{Helmholtz-Institut Mainz$,$ GSI Helmholtzzentrum für Schwerionenforschung GmbH$,$ 55128 Mainz$,$ Germany}
\affiliation{Johannes Gutenberg-University Mainz$,$ 55128 Mainz$,$ Germany}
\author{Dmitry Budker}
\affiliation{Helmholtz-Institut Mainz$,$ GSI Helmholtzzentrum für Schwerionenforschung GmbH$,$ 55128 Mainz$,$ Germany}
\affiliation{Johannes Gutenberg-University Mainz$,$ 55128 Mainz$,$ Germany}
\affiliation{Department of Physics$,$ University of California$,$ Berkeley$,$ California 94720-300$,$ USA}
\author{Ronald Ulbricht}
\email{ulbricht@mpip-mainz.mpg.de}
\affiliation{Max Planck Institute for Polymer Research$,$ Ackermannweg 10$,$ 55128 Mainz$,$ Germany}

%\date{21 December 2012}
%%%%%%%%%%%%%%%%%%%%%%%%%%%%%%%%%%%%%%%%%%%%%%%%%%
\date{\today}

\frenchspacing

\begin{abstract}
The negatively-charged nitrogen vacancy (NV$^{-}$) center in diamond is widely used for quantum sensing since the sensitivity of the spin triplet in the electronic ground state to external perturbations such as strain and electromagnetic fields make it an excellent probe for changes in these perturbations. The spin state can be measured through optically detected magnetic resonance (ODMR), which is most commonly achieved by detecting the photoluminescence (PL) after exciting the spin-triplet transition. Recently, methods have been proposed and demonstrated that use the absorption of the infrared singlet transition at 1042\,nm instead. These methods however require cryogenic temperatures or external cavities to enhance the absorption signal. Here, we report on our optimization efforts of the magnetometer sensitivity at room temperature and without cavities. We reach sensitivities of 18\,pT$/\sqrt{\mathrm{Hz}}$, surpassing previously reported values. We also report on a defect that is native to CVD-grown diamond and thus absent in HPHT diamond, the excitation of which impacts the measured singlet absorption signal.   
%We measure a noise floor of 28~pT/$\sqrt{\mathrm{Hz}}$ and magnetically insensitive noise floor of 18~pT/$\sqrt{\mathrm{Hz}}$. %We show that ODMR can be performed with a high dynamic range and discuss ways to further improve the performance of this method.
\end{abstract}
%%%%%%%%%%%%%%%%%%%%%%%%%%%%%%%%%%%%%%%%%%%%%%%%%%%
%\pacs{32.80.Lg, 32.30.Rj}

\maketitle

\section{Introduction}
The negatively-charged nitrogen-vacancy (NV$^{-}$) center in diamond is the most popular example of a new class of defects in solid-state crystals that promises novel applications in quantum information and metrology \cite{doherty2013nitrogen,strzalka2020qubit}, as well as magnetic \cite{barry2023sensitive}, electric \cite{electric1,block2021optically,electric3}, strain \cite{strain1,strain2} and temperature \cite{temp1,temp2} sensing.
Among them, due to promising improvements in both sensitivity and spatial resolution besides their robustness, wide operation temperature range, and small size, NV-based magnetometers have found numerous applications in condensed matter physics \cite{casola2018probing}, neuroscience and living systems biology \cite{barry2016optical}, nuclear magnetic resonance (NMR) \cite{wu2016diamond}, and Earth and planetary science \cite{glenn2017micrometer}. 

\begin{figure}[h]
    \centering
   % \captionsetup{skip=0pt} 
\includegraphics[width=\columnwidth]{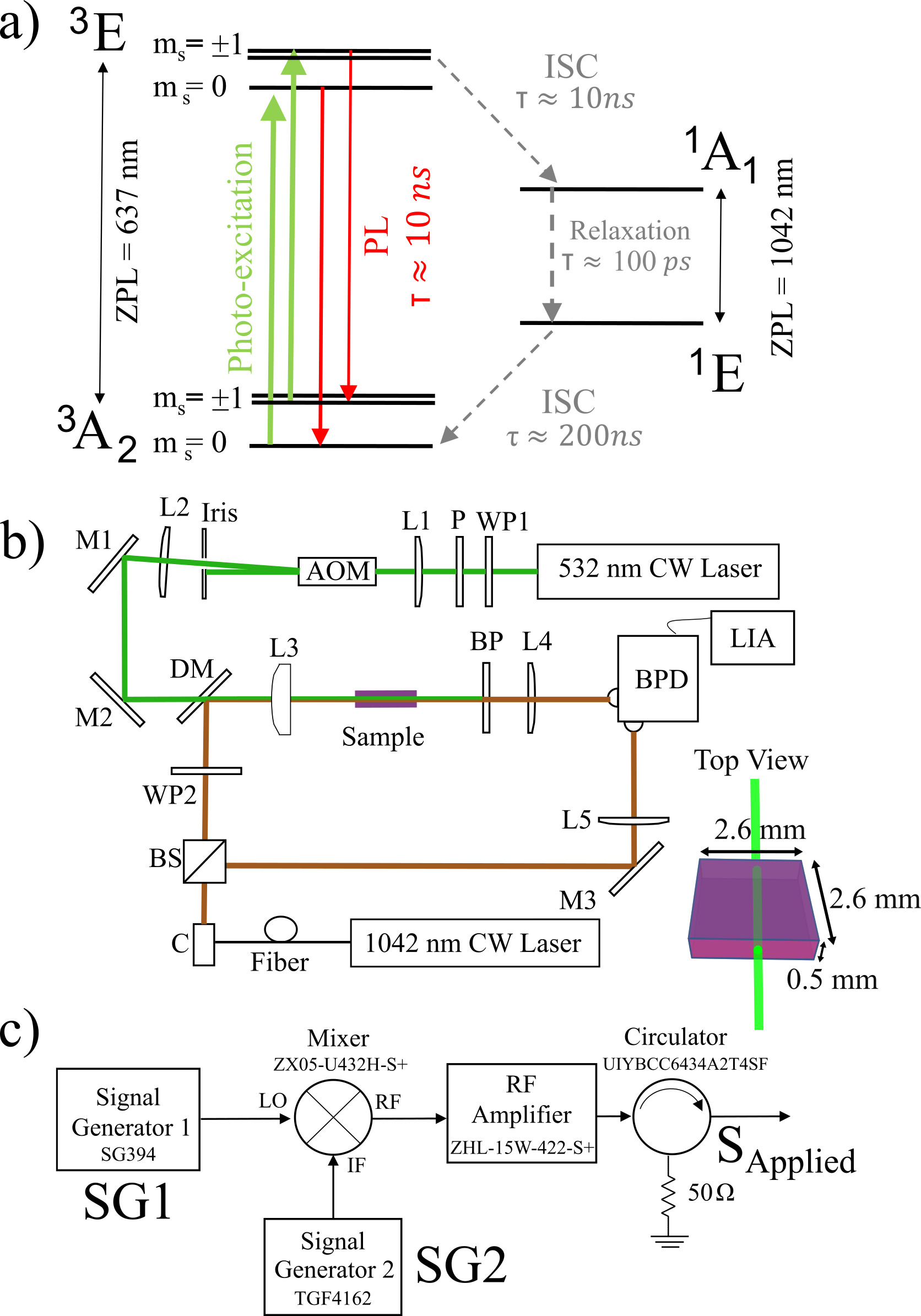}
    \caption{a) Electronic structure of negatively charged nitrogen vacancy (NV$^-$) centers with an optical pump and relaxation cycle and their lifetimes. b) Schematic of the implemented setup. WP: halfwave plate ($\lambda/2$), P: Polarizer, L: Lens, M: Mirror, DM: Dichroic mirror, C: Collimator, BS: Beam splitter, BP: Bandpass filter, BPD: Balanced Photodetector, LIA: Lock-in amplifier. c) MW signal generation system.}
    \label{fig:Fig1}
    %%%% ~237 words 
\end{figure}

The NV$^{-}$ center features a spin-triplet $^3A_2$ electronic ground state that can be manipulated with microwave (MW) radiation and read out through optical excitation as optically detected magnetic resonance (ODMR) \cite{hopper2018spin}.  
The utility of NV$^{-}$ for harnessing coherent spin dynamics relies on the reproducible polarization of this spin state through optical pumping, which is the result of a series of radiative and non-radiative electronic transitions that are shown in Fig.\,\ref{fig:Fig1}.a.
After photoexcitation into the spin-triplet excited state $^3E$, vibrational relaxation occurs within less than 100 fs \cite{ulbricht2018vibrational}. Further electronic relaxation takes place either via radiative recombination back to the $^3A_2$ state within 10 ns or via inter-system crossing (ISC) to the $^1A_1$ spin singlet. The latter transition is more probable from the spin projections states $m_s=\pm1$ states of $^3E$ compared to $m_s=0$ \cite{doherty2013nitrogen, SM}. Subsequently, the electron relaxes to the $^1E$ spin-singlet state within about 100 ps \cite{ulbricht2018excited}, from where it transitions back to the $^3A_2$ state via ISC within 200 ns at room temperature \cite{acosta2010optical,luu2024nitrogen}. Spin selectivity of this cycle not only leads to preferential population of $m_s=0$ in $^3A_2$ in the absence of external perturbations (spin polarization), but also allows $m_s$ to be read out optically (ODMR). 

MW radiation (or other controlled perturbations e.g. magnetic field) is used to manipulate the spin-projection states in $^3A_2$ which are eventually measured optically. The $m_s=\pm1$ states are degenerate at zero magnetic field and their energy is 2.87 GHz above $m_s=0$ (zero field splitting, ZFS). 
By applying a magnetic field, $m_s=+1$ and $m_s=-1$ split and this Zeeman splitting is proportional the projection of the magnetic field along the NV axis \cite{barry2020sensitivity}. The transition energy between \msz and \mspm states are given by $D\pm (g_e\mu_bB_{NV})/h$, where D is the ZFS, $g_e$ is NV electronic g factor, $\mu_B$ is Bohr magneton, $h$ is Planck's constant, and $B_{NV}$ is the magnetic field along the NV axis\,\cite{barry2020sensitivity}. 
Because of four axes of NV along [111],[1$\overline{1}\overline{1}$], [$\overline{1}\overline{1}1$], and [$\overline{1}1\overline{1}$], there will be separate splitting for each axis depending on the magnetic field's projection along them.
Features due to hyperfine coupling with surrounding nuclei, e.g. $^{14}\text{N}$, can be observed \cite{SM}. 

Commonly, ODMR is performed by detecting the PL of the $^3E \rightarrow$ $^3A_2$ transition, where a decrease in PL originates from a greater probability of relaxing via the ISC, which in turn is a result of a larger $m_s=\pm1$ population in the $^3A_2$ state. The $^1A_1 \rightarrow$ $^1E$ singlet transition could also be used for ODMR. In that case, $m_s=\pm1$ population will be correlated with its PL count \cite{ivady2021photoluminescence}. 
The PL yield of the singlet transition is however three to four orders of magnitude smaller than that of the triplet transition \cite{rogers2008infrared,acosta2010temperature,el2017optimised} due to dominant non-radiative recombination with a relaxation time of about 100 ps that is much faster than the radiative decay channel \cite{ulbricht2018excited}. The singlet PL is thus not an ideal signal for ODMR. The long lifetime of the $^1E$ state ($\approx 200$ ns) however suggests that absorption from $^1E \rightarrow$ $^1A_1$, instead of PL from $^1A_1 \rightarrow$ $^1E$, is a viable probe for ODMR. In this case, a probe beam with a photon energy that is resonant with the singlet transition would carry the absorption signal, enabling absorption ODMR \cite{acosta2010broadband, jensen2014cavity, chatzidrosos2017miniature, bougas2018possibility}. 

As compared to PL ODMR, absorption ODMR promises higher spin contrast due to the spin selectivity of the electronic states \cite{SM}. Since the absorption is coherently added to the probe beam as compared to the point source emission of PL, light collection is potentially more efficient in some probe geometries such as extended waveguide structures, fibers or cavities. Singlet absorption is also inherently insensitive to the presence of NV$^0$.  \\

The absorption coefficient of the singlet transition is however generally considered to be too low to be of practical use due to its low oscillator strength and broad linewidth at room temperature that leads to weak absorption values \cite{acosta2010broadband, ulbricht2018excited, munzhuber2020polarization}. Previous works that utilized the singlet absorption for ODMR have thus either operated at cryogenic temperatures \cite{acosta2010broadband}, where the transition linewidth is narrower and peak absorption thus larger, or placed the sample in a cavity to resonantly enhance absorption \cite{jensen2014cavity, chatzidrosos2017miniature}. Cavities were also used to enhance pump absorption and simultaneously perform ODMR through detection of the modulated pump transmission \cite{ahmadi2017pump, ahmadi2018nitrogen}.
\\
%%In this paper, we utilize the singlet absorption in a cavity-free configuration at room temperature with applying a continuous-wave (CW) MW to manipulate electron spins in the ground spin-triplet state. Contrary to common CW ODMR techniques, we employ a pulsed green pump for excitation and a CW IR laser at 1042 nm to readout. This results in enhanced IR absorption contrast compared to the CW pump scheme, which is discussed later. 
In this paper, we revisit the idea of utilizing the singlet ($^1E \rightarrow ^1A_1$) absorption at room temperature to optimize its performance without the aid of a cavity. We reach magnetic sensitivities of 18~pT/$\sqrt{\mathrm{Hz}}$, which is better than previous works that utilized cavities or worked at cryogenic temperatures. Avenues to further improve the sensitivity are discussed. We also found additional absorption contributions at the probe wavelength that originate from a defect native to CVD-grown diamond.   %The configuration of the sample used in this paper is 

\begin{figure*}[!t]
    \centering
\includegraphics[width=\textwidth]{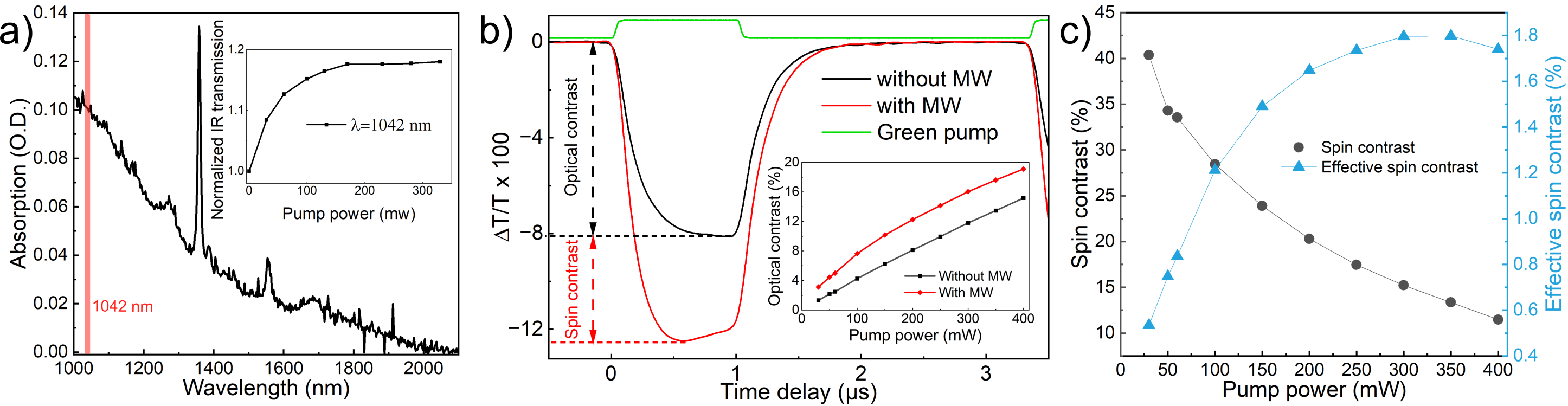}
    \caption{a) The absorption spectrum of the sample in the IR region. The normalized transmission of the IR beam at 1042\,nm for various CW pump powers is plotted in the inset. b) Differential transmission ($\Delta$T/T) at 1042\,nm with and without applied MW (red and black curves, respectively). The green plot shows when the pump is on and off. Inset: optical contrast for various pump powers. c) Spin contrast and effective spin contrast for various pump powers.}
    \label{fig:deltaT}
    %%% ~216 words
\end{figure*}

\section{experimental methods}
A CW laser at 532\,nm (Coherent Verdi) excites an ensemble of NV centers  ($^3A_2\rightarrow ^3E$) and a CW laser at 1042\,nm is used to probe the singlet transition $^1E\rightarrow^1A_1$. 
The 1042\,nm beam is generated with a tunable laser (TOPTICA DL pro) that is coupled to a polarization-maintaining single mode fiber. The outcoming beam of the fiber is collimated with a parabolic collimator and split into two paths, see Fig.\,\ref{fig:Fig1}.b. One beam is directly focused on one input of the balanced photodiode (PD, Femto Messtechnik) as a reference signal and the other passes through a halfwave plate ($\lambda/2$) that rotates the polarization of the IR beam to maximize the interaction with specific NV center orientations. It is deflected and focused with a spherical lens to a spot size of $\approx$ 24 $\mu$m diameter at the center of the sample. 

The output of the pump laser is attenuated with a halfwave plate and a polarizer (WP1 and P in Fig.\,\ref{fig:Fig1}.b), respectively.
The beam is focused on an acousto-optic modulator (AOM) (G\&H 3200-146, Crystal Technology) and the first order deflected beam is collimated with another lens (L2). The output beam is combined collinearly with the IR beam using a dichroic longpass mirror at 850\,nm (DM in Fig.\,\ref{fig:Fig1}.b), and focused with the same lens focusing the IR beam at the center of the sample, which is mounted on a 3-dimension translation stage. The spot size of the pump at the center of the sample is around 30\,$\mu$m and is assured to be larger than the IR beam over the entire sample. The Rayleigh length of the focused beams are calculated to be around 3\,mm.   
The sample is a $^{12}\text{C}$ enriched CVD grown diamond in [100] orientation, having substitutional nitrogen with a concentration of $\sim$10\,ppm \cite{SM}. The NV concentration (both negatively and neutrally charged) is about 4\,ppm. The sample size is 2.6$\times$2.6$\times$0.5\,mm$^3$ (polished on all six facets) and in order to have a better thermal dissipation, it is placed on a bigger diamond substrate of size 60$\times$60$\times$1\,mm$^3$. 
To increase IR absorption, the beams passes through the long side of the sample with a length of 2.6\,mm, thus increasing interaction without using a cavity. 
The transmitted beam is first passed through a 1040\,nm bandpass filter with a full-width-half-maximum of 10\,nm to block the pump beam and PL generated by the sample and focused on the second input of the balanced PD, whose output current is amplified with a built-in amplifier. 
The output of the amplifier is connected to a lock-in amplifier (LIA, Zurich Instruments HF2LI), which is used to demodulate the signal at a specific frequency that is explained later. 

A bias permanent magnet is placed oriented along [110], which has equal field strength on two NV axes ([111] and [$\overline{1}\overline{1}1$]) and is orthogonal to the other two ([1$\overline{1}\overline{1}$] and [$\overline{1}1\overline{1}$]). The strength of the magnetic field at the sample distance is around  1\,mT. %0.9
The MW signal is applied with a wire placed on top of the sample with no physical contact. %The details of the MW delivery system whose schematic is shown in Fig. \ref{fig:Fig1}.c is explained later. 
 Due to $^{14}\text{N}$ nuclear spin in this sample, each transition from $m_s=0\leftrightarrow m_s=\pm1$ has three hyperfine components. Addressing all of these hyperfine splittings simultaneously increases the ODMR contrast \cite{SM}. Having the MW resonant with both $m_s=0\leftrightarrow m_s=\pm1$ also increases the ODMR contrast \cite{SM}. Therefore, a set of radiofrequency components is implemented to address all the hyperfine and both electronic state transitions, see Fig.\,\ref{fig:Fig1}.c. 
 The MW signal is generated at the frequency of $f_{SG_1}$, which is set at the zero-field splitting of the $^3A_2$ ( $f_{SG_1}=f_{ZFS}$). A second signal generator producing a customized sine wave at three frequencies is mixed with the output signal of the MW signal generator producing six components around the ZFS. The frequencies are adjusted to drive all hyperfine transitions for both spin state transitions. This signal is then amplified and applied to the wire placed on the sample, see SM for more details.

\section{Results and discussions}
In order to measure the pump-induced absorption of the IR probe by the singlet transition, we measured the transmitted beam intensity while the sample is pumped and unpumped. It is expected that the transmitted IR intensity decreases as the pump intensity increases due to absorption by the singlet transition as the $^1$E level is populated  \cite{chatzidrosos2017miniature,jensen2014cavity,younesi2022broadband}. 
We however also noticed a competing process through which, upon increasing the pump power, the transmission of the IR probe increases, see inset in Fig.\,\ref{fig:deltaT}.a. This is presumably due to the absorption of other defects in the sample that absorb less (i.e. bleach) when excited by the pump.  
Interestingly, this additional competing process occurs only in our CVD-grown samples and not in high-pressure high-temperature (HPHT) diamond. IR absorption measurements reveal a pronounced ZPL at 1358\,nm in the CVD samples that is accompanied by a broad phonon sideband stretching to the singlet absorption region at 1042\,nm, see Fig.\,\ref{fig:deltaT}.a. This defect is presumed to be related to hydrogen defects and predominantly occurs in CVD diamond \cite{fuchs1995hydrogen,zaitsev2013optical}. It is conceivable that the inevitable co-excitation of this additional defect during a magnetometry measurement has a detrimental effect on the NV spin relaxation properties, which could potentially favor HPHT diamond over CVD diamond in certain scenarios. It would thus be interesting for future studies to investigate this further. 

In order to quantify the absorption of the IR beam due to singlet-level absorption only, i.e. to correct for the pump-induced bleaching of the 1358\,nm defect absorption, a pulsed pump transient absorption technique was used \cite{younesi2022broadband}. Here, the pump power is modulated with the AOM at 300 kHz with a duty cycle of 30\% and the probe transmission of the sample is recorded on the oscilloscope after a transimpedence amplification stage of the photodiode output current. The differential transmission plot of the sample at 1042\,nm with and without applied MW is shown in Fig.\,\ref{fig:deltaT}.b. We define \textit{optical contrast} as the maximum differential transmission modulation $\Delta$T/T, which as expected increases linearly with pump power (shown in the inset of Fig.\,\ref{fig:deltaT}.b). The optical contrast without MW ($C_{noMW}$), which in Fig.2.b is 8\%, whereas maximum $\Delta$T/T is about 12\% with MW ($C_{MW}$). The contrast between these two values then determines the \textit{spin contrast}, as defined by $(C_{MW}-C_{noMW})/(C_{MW}+C_{noMW})$ \cite{SM}. In PL ODMR, all detected NV-PL emission (not counting possible NV$^0$ emission) contribute to the ODMR signal, rendering the optical contrast to essentially 100\%. However, this is not the case for absorption ODMR, where the signal will be determined by the product between optical and spin contrast, which we define as the \textit{effective spin contrast} $C_{CW}$. The power dependence of both spin contrast and effective spin contrast is shown in Fig.\,\ref{fig:deltaT}.c. Note that this pulsed pump scheme is only used to measure the transient absorption signal, whereas all magnetometry measurements are based on CW pumping.

%As explained in the previous sections, the pump-induced absorption is obtained at the output of the LIA.
%In the presence of a magnetic field, the transition energy between \msz and \mspm states are given by $D\pm (g_e\mu_bB_{NV})/h$, where D is the ZFS, $g_e$ is NV electronic g factor, $\mu_B$ is Bohr magneton, $h$ is Planck's constant, and $B_{NV}$ is the magnetic field along NV axis. Because of four orientations of NV along [111],[1$\overline{1}\overline{1}$], [$\overline{1}\overline{1}1$], and [$\overline{1}1\overline{1}$], there will be separate splitting for each axis depending on the magnetic field's projection along them. 
%Here, a permanent ring magnet is applied along (110) that results in the same projection on all NV axes, and their splitting overlap with each other leading to higher contrast in the ODMR signal. The normalized absorption change due to applying the MW signal is shown in Fig. \ref{fig:CWODMR}. (we'd better apply it along two axes because there's a projection coefficient division for the magnetic field conversion ($\cos(54\degree)$), instead in two overlap, we will have $\cos(35\degree)$ at (110) orientation).
\begin{figure}[t]  
\raggedright
\includegraphics[width=0.45\textwidth]{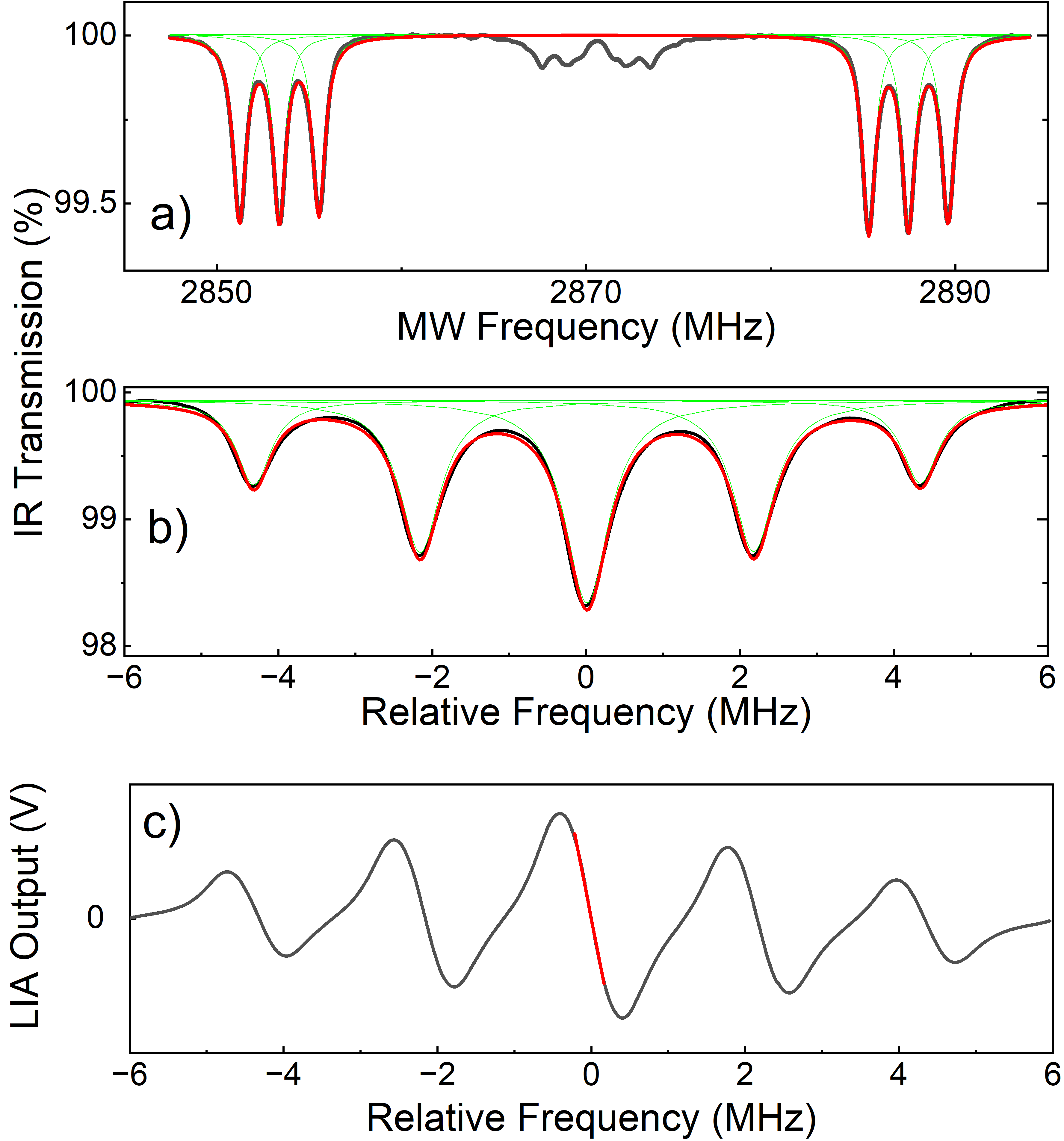}
    \caption{ODMR plots measured by IR transmission while MW field is scanned. a) A single MW source is used (SG1). b) The mixed MW signal addressing all hyperfine splittings for $m_s=0\leftrightarrow m_s=\pm1$. c) The demodulated output of LIA when the MW field is modulated. A bias magnetic field along [110] is applied with a strength of around 1\,mT at the sample.}
    \label{fig:CWODMR}
     %%% ~200 words
\end{figure}
To perform absorption ODMR, we CW-pump the sample and measure the transmission of the IR beam while scanning the MW signal across the resonances, shown in Fig.\,\ref{fig:CWODMR}.a for a pump power of 200 mW and an applied bias field. The peaks are fitted with six Lorentzian functions \cite{SM}. 
In this case, only the output of signal generator 1 (at $f_{SG_1}$) is connected to the RF amplifier (bypassing the mixer, shown in Fig.\,\ref{fig:Fig1}.c). 
In Fig.\,\ref{fig:CWODMR}.b, the mixed MW signal is applied to address all hyperfine and both $m_s=0\leftrightarrow m_s=\pm1$. Here, $f_{SG_1}$ is fixed at ZFS, 2.870 GHz and $f_{SG_2}$ is scanned around 17 MHz. In this case, the central peak has the highest effective spin contrast $C_{CW}=1.6 \%$ with a full-width-half-maximum (FWHM, $\Delta\nu$) of 700 kHz.

To measure the magnetic field, the MW signal is modulated around the central peak frequency $f_{SG_2}$ with a modulation frequency of $f_{mod}=5.6 ~\mathrm{kHz}$ and a deviation amplitude of $f_{dev}=330 ~\mathrm{kHz}$ sweeping the signal generator frequency according to: $f_{dev}\sin(2\pi f_{mod}t)$. By demodulating the transmitted IR signal using the LIA at the frequency of $f_{mod}$, we obtain a set of dispersive resonances shown in Fig.\,\ref{fig:CWODMR}.c. 
At the zero-crossing of this curve, the LIA output signal is linear to the difference between the central MW and the magnetic resonance frequency, $S_{LIA}\sim \alpha(f_c-f_{res})$ for small deviation of $|f_c-f_{res}|\ll\Delta\nu/2$ with the coefficient $\alpha\propto C/\Delta\nu$. This linear response at zero-crossing is fitted with the red line in Fig.\,\ref{fig:CWODMR}.c. 
Small changes in magnetic field are proportional to the LIA output signal and the magnetic field is then calculated by the following equation: $\Delta B_{NV}=S_{LIA}~h/(g_e\mu_B\alpha\cos(\theta))$, where $\theta$ is the angle between the measured magnetic field (along [110]) and two NV axes with equal field projection (along [111] and [$\overline{1}\overline{1}1$]), which is equal to $35.3\degree$ \cite{SM}. It is derived from the equation of the Zeeman splitting of the spin states discussed earlier. 

%\begin{align}
%f_{app}&=f_{MW_S} \pm (f_c + f_{dev}\sin(2\pi f_{mod}t) + \notag \\  &\quad f_{MW_S} \pm ((f_c+2.16 \text{ MHz}) +  f_{dev}\sin(2\pi f_{mod}t))+ \notag   \\ &\quad f_{MW_S} \pm ((f_c-2.16 \text{ MHz}) + f_{dev}\sin(2\pi f_{mod}t))
%\end{align}

%\begin{align}
%S_{applied}&=\sin(f_{MW_S} \pm (f_c + f_{dev}\sin(2\pi f_{mod}t)) + \notag \\  &\quad \sin(f_{MW_S} \pm ((f_c+2.16 \text{ MHz}) + \notag \\  &\quad  f_{dev}\sin(2\pi f_{mod}t)))+ \notag \\ &\quad \sin(f_{MW_S} \pm ((f_c-2.16 \text{ MHz}) + \notag   \\ &\quad f_{dev}\sin(2\pi f_{mod}t)))
%\end{align}

\begin{figure}[t]
    \centering
\includegraphics[width=0.48\textwidth]{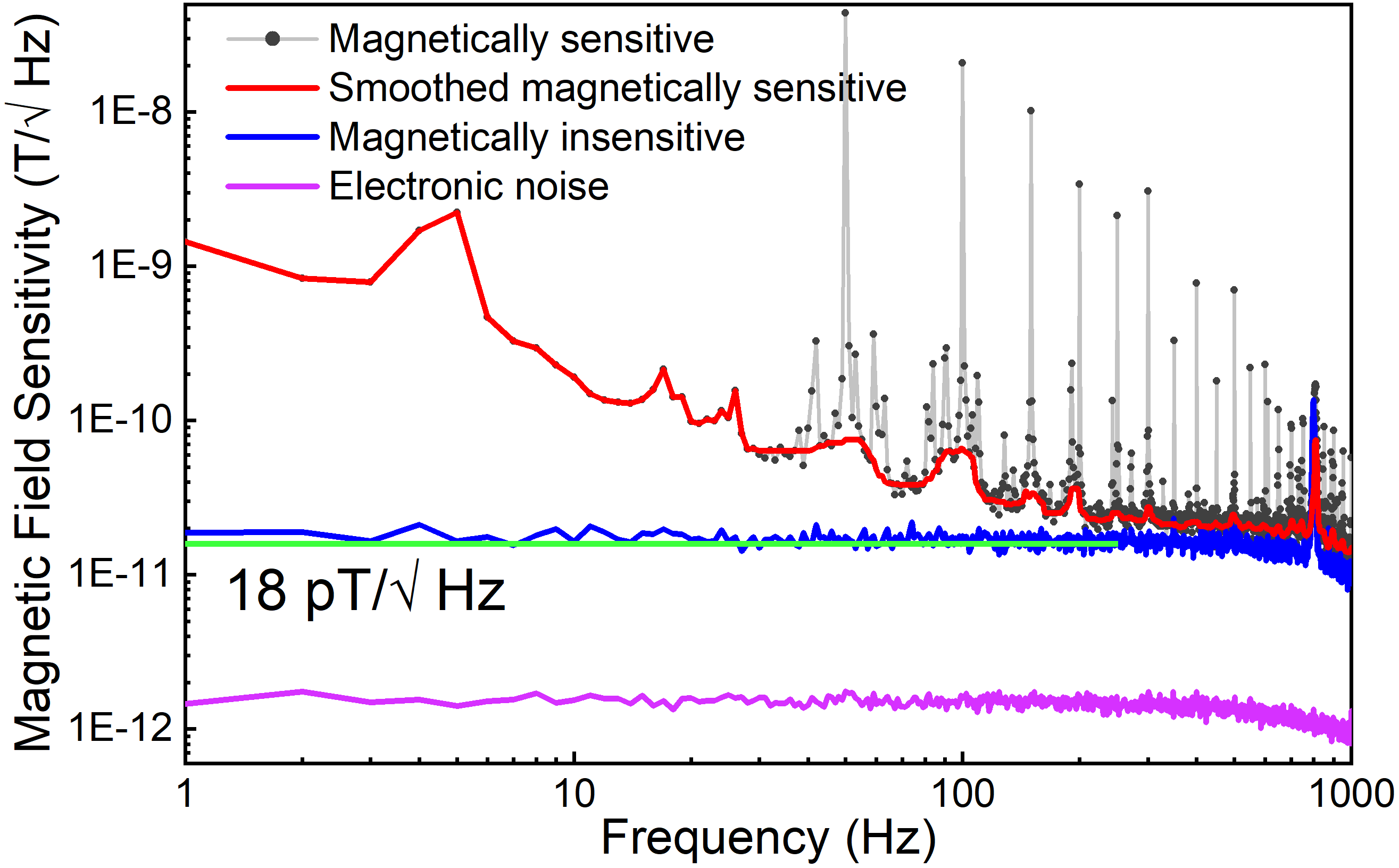}
\caption{Relative amplitude spectra of the sensor to characterize the magnetic field sensitivity. It is plotted for magnetically sensitive case (black and smoothed curve in red), magnetically insensitive, and electronic noise. The sensitivity noise floor of  18\,pT/$\sqrt{\mathrm{Hz}}$ is also illustrated in green line.}
    \label{fig:Sens}
    %%%~125 WORDS
\end{figure}

The magnetic field sensitivity, measured in an unshielded environment in the laboratory, is shown in Fig.\,\ref{fig:Sens}. It is calculated using the linear spectral density (LSD) of the recorded signal (magnetic field) in time \cite{SM}. The data plotted in Fig.\,\ref{fig:Sens} is the 60 times average of the LSD of 1-second segments recorded data (see \cite{SM} for further details). 
%%The In order to calculate the sensitivity, the LIA output is first recorded for 60 seconds and is converted to the magnetic field with the conversion ratio. Then, the data is split to 60 1-second long measurements and the power spectral density for each 1-second segment is calculated using the Fourier transform and by having the square root of that. Finally, all of the spectral density curves are averaged together, which gives the sensitivity graph. 
The sensitivity is plotted for three cases: magnetically sensitive, magnetically insensitive, and electronic noise. In order to obtain the magnetically sensitive one, the MW is applied at the resonance, and all present magnetic fields including the 50\, Hz power line and its higher harmonics are visible in the measurement. Because of magnetic laboratory interference in the magnetically sensitive curve, it is smoothed with a moving median filter for better illustration (red plot). For magnetically insensitive measurement, the MW signal is applied far from the resonances, having no effect on the spin of the ground state electrons. In this case, all non-magnetic noise sources including laser fluctuations are contributing which leads to a sensitivity noise floor of 18\,pT/$\sqrt{\mathrm{Hz}}$, representing the lowest value reported so far using IR absorption NV-based magnetometers. Finally, the electronic noise is measured by blocking all the light sources, reaching the noise floor of 1.5~pT/$\sqrt{\mathrm{Hz}}$. The bandwidth (BW) of the magnetic field measurement is defined by the lowpass filter of the LIA, which is set to 1 kHz. The magnetic field sensitivity of the setup is further measured using a test magnetic field, see SM \cite{SM}. Based on the measured parameters, the shot-noise sensitivity is calculated to be around 5~pT/$\sqrt{\mathrm{Hz}}$ \cite{SM}.  

\section{Conclusion}
In this report, we discussed a magnetometer that employs ODMR detected via absorption of the IR singlet transition in NV$^{-}$ centers at room temperature and without cavity enhancement. The magnetometer reaches a sensitivity of 18\,pT/$\sqrt{\mathrm{Hz}}$ from DC to 900\,Hz, the lowest value reported so far using absorption ODMR. An additional diamond defect was identified in CVD-grown samples that absorbs in the IR probing range, which may be a limiting factor for further development due to the additional probe-light absorption. Substituting CVD diamond with HPHT diamond could therefore be a sensible choice when building IR absorption ODMR magnetometers.

There are several avenues for pushing the sensitivity further. The most promising is to increase the probe absorption, and thus the optical contrast, in a multi-path (e.g. cavity) beam propagation geometry. Pump-power stabilization and introducing magnetic shielding should reduce the influence of technical noise, pushing the sensitivity closer to the shot-noise limit, which in turn could be decreased even further by increasing the IR probe power.
\\

%We introduced a single-path (without cavity enhancement) room temperature magnetometry setup using IR absorption in NV centers, reaching a sensitivity of around 18\,pT/$\sqrt{\mathrm{Hz}}$. This sensitivity is the lowest value reported using IR absorption NV based magnetometers. 
%Having higher sensitivity than the estimated shot-noise limit is promising for further improvements in the sensitivity. 
%This technique is more advantageous where the PL collection efficiency cannot be extended easily, e.g. in endoscope applications. Further improvements in the sensitivity can be achieved using multi-path IR propagation over the sample or putting the sample in a cavity. 

%Additionally, since sensitivity and shot-noise limit are determined by the IR laser power and not the PL, it can even be further suppressed reaching better sensitivity by increasing the IR power.
%However, an additional defect was identified in CVD-grown samples, which may be a limiting factor for further sensitivity suppression due to IR light absorption in the sample. Consequently, substituting with an HPHT sample with the same parameters is a more promising avenue for further developments. 

%Another approach for making the setup more practical is to employ another light source, e.g. high-power LED, instead of a 1042\,nm laser which makes the setup less complicated and more affordable. 

\noindent \textbf{Acknowledgements}

We acknowledge Marc-Jan van Zadel and Florian Gericke for technical support and Mahsa Ebrahimi for help with preliminary experiments. We acknowledge funding by the Max-Planck Society. 
We also acknowledge support from the European Commission’s Horizon Europe Framework Program under the Research and Innovation Action MUQUABIS GA no. 101070546, by the German DFG, Project SFB 1552 "Defekte und Defektkontrolle in weicher Materie” and funding by the Carl-Zeiss-Stiftung (HYMMS P2022-03-044), also the German Federal Ministry of Education and Research (BMBF) within the Quantumtechnologien program through the DIAQNOS project (project no. 13N16455).

\bibliographystyle{unsrt}
\bibliography{biblography.bib}

\end{document}